\documentclass[aps,twocolumn]{revtex4}
\usepackage{graphicx}
\usepackage{dcolumn}
\usepackage{bm}
\usepackage{amsmath}

\begin{document}

\title{Enhanced laser cooling of rubidium atoms in two-frequency diffuse lights}
\author{Wen-Zhuo Zhang$^{1,2}$}
\author{Hua-Dong Cheng$^1$}
\author{Ling Xiao$^1$}
\author{Liang Liu$^{1,*}$}
\author{Yu-Zhu Wang$^1$}

\affiliation{$^1$Key Laboratory of Quantum Optics, Shanghai
Institute of Optics and Fine Mechanics, Chinese Academy of Sciences,
Shanghai 201800, P. R. China.} \affiliation{$^2$Graduate University
of the Chinese Academy of Sciences, Beijing 100039, P. R. China,}
\affiliation{$*$Corresponding author: liang.liu@siom.ac.cn}

\date{\today}

\begin{abstract}
In this paper we describe an experiment of efficient cooling of
$^{87}$Rb atoms in two-frequency diffuse laser lights.  Compared
with single frequency diffuse light, two-frequency diffuse lights
have wider velocity capture range and thus can cool more atoms. In
our experiment, the maximum number of cooled atoms can reach up to
$3.9\times10^9$. Such a result is quite useful in building a compact cold atom clock.
\end{abstract}

\maketitle

\noindent

Cooling of atoms in diffuse laser lights is an all-optical laser
cooling method with a wide velocity capture range compared with
optical molasses. It was first applied in slowing and cooling of
atomic beams \cite{Ketterle1992,Batelaan1994,Chen1994,Wang1995}.
Later, three-dimensional cooling of Cs atoms from background vapor
in one-frequency diffuse laser lights was also realized
\cite{HORACE_OL_2001,Cheng_PRA_2008}. This technique leads to a
development of compact cold-atom clock (The HORACE)
\cite{HORACE_IEEE_2001,HORACE_IEEE_2004,HORACE_IEEE_2005}, which
gives an appealing stability at $5.5\times 10^{-13}$ at 1 sec.
\cite{HORACE_IEEE_2007}.

There are still some rooms for the improvement of the performance of
the HORACE. For example, increasing the number of detected atoms can
reduce both the quantum projection noise and the detection noise
\cite{HORACE_IEEE_2005}. This is especially useful for rubidium
atoms because collision shift of cold rubidium atoms is much smaller
than cesium \cite{Sortais_IEEE_2000}. In this paper, we present an
experiment of two-frequency diffuse light cooling of rubidium atoms
directly from background vapor, with which more cold atoms are
captured than in single-frequency diffuse light used in previous
experiments \cite{HORACE_OL_2001,Cheng_PRA_2008}.

As well known, the radiation force for a monochromatic laser on a
two-level atom is
\begin{equation}\label{single_force}
F=\hbar k{\Gamma\over 2} {s \over 1+s+(2\Delta/\Gamma )^2}
\end{equation}
where, $\Gamma =2\pi\cdot6.066$MHz is the decay rate of the excited
state, $s$ is the on-resonance saturation parameter, $\Delta $ is
the detuning between laser and atom. In diffuse light, for an atom
moving at velocity $\vec v$ and interacting with a laser beam with
an angle $\theta $ with respect to $\vec v$, the detuning is
$\Delta=\omega -\omega _a-\vec k\cdot\vec v=\Delta+kv\cos\theta$,
$\omega$ and $\omega_a$ are the frequency of laser and atomic
transition respectively. We can easily have the force on a two-level
atom in a pair of oppositely propagating light beams from
Eq.(\ref{single_force}), one with an angle $\theta $ with respect to
the opposite direction of atomic velocity, and the other with an
angle $\theta $ with respect to the same direction of atomic
velocity:
\begin{equation}\label{diffuse_force}
\begin{split}
F=&-\hbar k{\Gamma\over 2}\frac{s
\cos\theta}{1+s+4(\Delta+kv\cos\theta)^2/\Gamma ^2}\\
&+\hbar k{\Gamma\over
2}\frac{s\cos\theta}{1+s+4(\Delta-kv\cos\theta)^2/\Gamma ^2}
\end{split}
\end{equation}

In diffuse light, the on-resonant light dominates the interaction
process between the atom and the diffuse light. Thus
Eq.(\ref{diffuse_force}) is a good approximation for considering
atomic motion in the diffuse light at the condition
\begin{equation}\label{resonant_condition}
\Delta+kv\cos\theta=0
\end{equation}
Obviously, for cooling, $\Delta$ must be negative. The lowest
velocity for which the Eq.(\ref{resonant_condition}) can be
satisfied is $kv_{\min}=|\Delta|$ when $\theta=0$. On the other
hand, from Eq.(\ref{diffuse_force}), the radiation force drops as
$\theta$ increases while the Eq.(\ref{resonant_condition}) is still
satisfied. For example, the radiation force drops to half of the
maximum when $\cos\theta=1/2$, which gives $kv_{\max}=2|\Delta|$.
Thus we have an efficient cooling range of velocity for a two-level
atom in diffuse light $|\Delta|\leq kv \leq 2|\Delta|$. If we choose
$\Delta=-\Gamma$, as typical optical molasses does, the capture
velocity is $2\Gamma/k$, which is in the same order with typical
optical molasses.

In order to increase the velocity capture range, the detuning
$|\Delta|$ must be large, but large detuning leads to large final
velocity of cooled atoms. A simple way to solve this problem is the
use of multiple diffuse light frequencies as used by Ketterle {\it
et al.} in cooling of an atomic beam \cite{Ketterle1992}. Atoms with
high velocity are cooled by large-detuned light, and those with low
velocity by small-detuned light \cite{Zhang_AOS_2007}.
\begin{figure}
\centerline{\includegraphics[width=3.in]{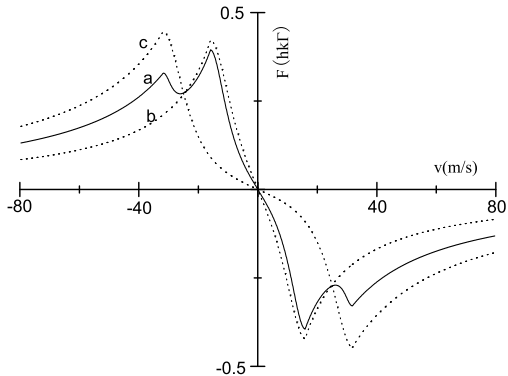}}
\caption{\label{force}Force of diffuse lights vs atomic velocity:
(a) two-frequency lasers with $\Delta_1$=$-3\Gamma$,
$\Delta_2$=$-6\Gamma$, $s_1=s_2=5$; (b) one-frequency laser with \
$\Delta$=$-3\Gamma$, $s=10$; (c) one-frequency laser with \
$\Delta$=$-6\Gamma$, $s=10$.}
\end{figure}
In fact, the cooling of an atomic beam is more like the
deceleration of atoms, while the cooling of atoms from background
vapor is more like the accumulation of cold atoms besides cooling.
Thus cooling from background vapor is more attractive because it
has a simple structure and the cooled atoms is ``localized" in a
region where many experiments can be done. We calculated the force
on a two-level atom in two-frequency diffuse lights as
\begin{equation}\label{two_freq_force}
\begin{split}
F=&-\hbar k {\Gamma \over 2}\sum_{n=1}^2\frac{s_n\cos\theta_n
\alpha}{1+4(\Delta_n+kv\cos\theta_n)^2/\Gamma^2}\\
&+\hbar k {\Gamma \over 2}\sum_{n=1}^2\frac{s_n\cos\theta_n\beta}
{1+4(\Delta_n-kv\cos\theta_n)^2/\Gamma^2}
\end{split}
\end{equation}
where
\begin{equation}
\alpha =1/[1+\sum_{n=1}^2{s_n \over
1+4(\Delta_n+kv\cos\theta_n)^2/\Gamma^2}]
\end{equation}
and
\begin{equation}
\beta=1/[1+\sum_{n=1}^2{s_n \over
1+4(\Delta_n-kv\cos\theta_n)^2/\Gamma^2}].
\end{equation}
Here $s_1,\Delta_1$ and $s_2,\Delta_2$ are the on-resonant
saturation parameters and detunings for laser 1 and laser 2
respectively, and we assume that $|\Delta_2|>|\Delta_1|$.
Eq.(\ref{two_freq_force}) is valid for both $kv\geq|\Delta_1|$ and
$kv<|\Delta_1|$. For atoms with velocity $v<|\Delta_1|/k$, the force
is similar to the optical molasses \cite{Lett_JOSA_1989}. Fig.
\ref{force} gives a plot of force on a two-level atom in diffuse
lights, which is a combination of different velocity ranges. From
the figure, we can see that for small detuning, the velocity capture
range is small, while for large detuning, the force for small
velocity is small, and thus it can not cool atoms to very low
velocity. Only with the combination of two lasers with appropriate
frequencies, the lights can cool atoms over wide velocity range to
very low temperature, as shown in the solid line of Fig.
\ref{force}.
\begin{figure}
\centerline{\includegraphics[width=3.in]{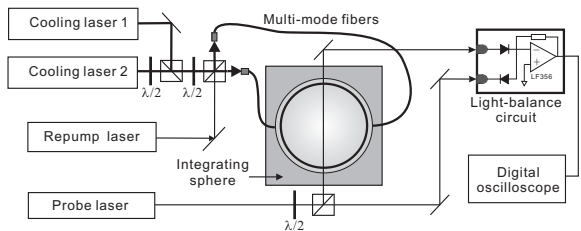}}
\caption{\label{setup} Setup of the two-frequency diffuse laser
cooling.}
\end{figure}

Fig. \ref{setup} shows the experimental setup for two-frequency
diffuse light cooling of atoms in an integrating sphere, similar to
the one described in Ref.\cite{Cheng_PRA_2008}. Two cooling lasers
are locked at the transition $5^2$S$_{1/2},F=2\rightarrow
5^2$P$_{3/2},F'=3$ of $^{87}$Rb and the frequencies are shifted by
Acousto-Optic Modulators (AOMs). A weak repumping laser with total
power of 4.7 mW, mixed with the cooling lasers, is locked to the
transition $5^2$S$_{1/2},F=1\rightarrow 5^2$P$_{3/2},F'=2$. This
repumping laser is used  to pump the population trapped in the
$5^2$S$_{1/2},F=1$ back to the cooling state $5^2$S$_{1/2},F=2$. All
lasers are injected into an integrating sphere through two
multi-mode fibers. Diffuse lights are formed inside the sphere when
the lasers are multi-reflected by its inner surface whose
reflectivity is 98\% at $780nm$ wavelength. Inside the integrating
sphere, a glass cell with an inner diameter of 43 mm, mounted on a
vacuum system of $~10^{-7}$ Pa, is connected to a rubidium reservoir
which is kept at room temperature and supplies the cell the rubidium
vapor.

A very weak probe laser with power of 1 $\mu$W, locked at the
transition $5^2$S$_{1/2},F=2\rightarrow 5^2$P$_{3/2},F'=3$, is
placed vertically. When cold atoms are accumulated in the glass cell
during the cooling, an absorption of the probe beam is recorded.
Since the cold atoms are saturated by cooling lasers, the recorded
signal does not reflect the real absorption of the probe beam by the
cold atoms \cite{HORACE_OL_2001}. In order to measure the
undisturbed properties of the cold atoms, we need to switch off all
cooling lasers when the absorption signal become stable.
\cite{Cheng_PRA_2008}.
\begin{figure}
\centerline{\includegraphics[width=3.in]{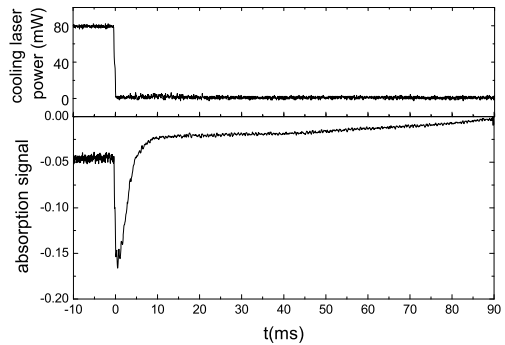}}
\caption{\label{absorption}Absorption of probe laser by cold atoms
vs time. The peak value of absorption signal when cooling laser is
switched off is proportional to the real number of cold atoms.
$\Delta_1=-3\Gamma$, $\Delta_2=-5\Gamma$ and $P_1=P_2=40mW$.}
\end{figure}

Fig. \ref{absorption} shows a typical absorption signal vs time.
Here $\Delta_1=-3\Gamma$ and $\Delta_2=-5\Gamma$. Total power of the
two-frequency diffuse laser lights in the integrating sphere is
$80mW$. When no cooling laser lights are injected into the
integrating sphere, the signal gives the absorption of the probe
beam by background vapor (which is fairly small compared with
absorption signal of cold atoms). After cooling lasers are turned
on, the absorption of the probe beam by cold atoms are gradually
increased until it is stable. The loss of cold atoms comes mainly
from the drop by gravity after the atoms are cooled, and once the
accumulation and loss are balanced, the number of the cold atoms in
the cell is stable (absorption signal in FIG.2 when $t<0$). After
the absorption signal is stable, we suddenly switch off the cooling
lasers (at $t=0$ in FIG.2), but keep the repumping laser on. The
absorption of the probe beam by cold atoms is suddenly increased
because they are no longer saturated by the cooling lights. The peak
does give the real absorption signal of the probe beam by cold
atoms. The quickly decreased absorption is due to the decreasing of
the cold atom's density, which is resulted from the momentum
diffusion of cold atoms excited by the probe beam
\cite{Cheng_PRA_2008}.
\begin{figure}
\centerline{\includegraphics[width=3.in]{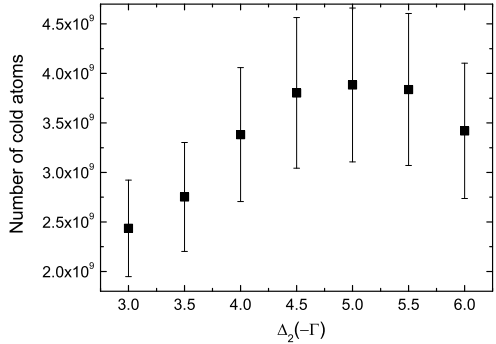}}
\caption{\label{number of_cold atoms} Number of cold atoms vs
$\Delta_2$ with fixed $\Delta_1=-3\Gamma$. Power of both lasers
injected into the integrating sphere is $40mW$.}
\end{figure}

The absorption peak in Fig.\ref{absorption} represents a real
absorption of the probe beam for cold atoms in the glass cell, from
which we can determine the number of cold atoms in the glass cell.
Fig.\ref{number of_cold atoms} gives number of cooled atoms by
two-frequency diffuse lights vs detuning $\Delta_2$ of laser 2 with
fixed detuning $\Delta_1=-3\Gamma$ of laser 1. The two-frequency
diffuse light cooling is equivalent to single-frequency diffuse
light cooling when $\Delta_2=\Delta_1$. In Ref.
\cite{Cheng_PRA_2008}, we already proved that for single frequency
diffuse cooling, the maximum number of cold atoms can be obtained
when the detuning is around $-3\Gamma$. In Fig. \ref{number of_cold
atoms}, when $\Delta_2=\Delta_1=-3\Gamma$ the maximum number of cold
atoms in single frequency diffuse light is obtained. As $\Delta_2$
increases, more atoms are cooled and accumulated, and maximum number
of cold atoms reaches $3.9\times 10^9$ when $\Delta_2$=$-5\Gamma$.
This is a clear evidence of the increasing capture power of cold
atoms in two-frequency diffuse lights.

Several factors limit the further increasing of cold atoms in the
two-frequency diffuse lights. The number of the captured atoms
depends on the number of atoms over the capture range of
two-frequency diffuse light in the background vapor. After those
atoms are cooled, the vapor atoms again become equilibrium through
collision, and atoms over the capture range are produced again.
The cycle keeps the cooling and capturing of atoms continuous
until it is balanced to the loss of cold atoms discussed
previously. Increasing the background vapor pressure can increase
both the number of atoms over the capture range and collision
between atoms, and thus can increase the captured cold atoms.
Certainly, increasing the vapor pressure also increases the
collision between hot atoms and cold atoms, which is a damage to
the cold atoms.

In conclusion, we have demonstrated the increasing capture power
of atoms in two-frequency diffuse lights, and discussed the
possibility of further improvement. The increased number of cold
atoms is useful for the improvement of an atomic clock using cold
atoms in an integrating sphere.

This work is supported by the National Nature Science Foundation
of China under Grant No. 10604057 and National High-Tech Programme
under Grant No. 2006AA12Z311.

\end{document}